\documentclass[onecolumn,floatfix,showpacs,aps,nofootinbib, showkeys]{revtex4-2}

\usepackage[dvips]{epsfig}
\usepackage[english]{babel}

\usepackage{bbm}
\usepackage{verbatim}
\usepackage{array}
\usepackage{amsmath}
\usepackage{multirow}
\usepackage{amsfonts}
\usepackage{amssymb}
\usepackage{hyperref}
\usepackage{bm,upgreek}
\usepackage{leading}
\usepackage[dvipsnames]{xcolor}
\hypersetup{colorlinks=true,linkbordercolor=Blue,linkcolor=Blue, citecolor=Blue}
\usepackage{subeqnarray}
\usepackage{setspace}
\usepackage{indentfirst} 

\newcommand{\boldp}{(p^{\mu})}
\newcommand{\boldo}{(k^{\mu})}
\usepackage{gensymb}
\usepackage{bbm}

\AtBeginDocument{
    \hypersetup{colorlinks,urlcolor=Blue}
}

\begin{document}

\title{Spinorial structures, discrete symmetries and some consequences}

\author{C. H. Coronado Villalobos$^{1}$} \email{ccoronado@autonoma.edu.pe}
\author{R. J. Bueno Rogerio$^{2}$} \email{rodolforogerio@gmail.com}
\author{A. R. Aguirre$^{2}$} \email{alexis.roaaguirre@unifei.edu.br}
\affiliation{$^{1}$Universidad Aut\'onoma del Per\'u, 
Panamericana Sur Km 16.3, Villa el Salvador, Lima, Per\'u,}
\affiliation{$^{2}$ Instituto de F\'isica e Qu\'imica, Universidade Federal de Itajub\'a (UNIFEI),\\Av. BPS 1303, CEP 37500-903, Itajub\'a, MG, Brasil.}


\begin{abstract}
\noindent{\textbf{Abstract.}}
In this paper  we discuss fundamental aspects related to the helicity and dynamics of the spin-$1/2$ fermions encompassed within the very well-known Lounesto's classification. More specifically,  we investigate how the bi-spinorial structures behave under discrete symmetries, as well as analyse some consequences on the spinors dynamics. In addition, we find an interesting relation between  the spinor helicity and the Lounesto spinor classification.
\end{abstract}


\maketitle

\section{Introduction}\label{intro}

It is currently understood that spinors can be characterized as Dirac spinors, which describe for instance electrons and positrons; Majorana describe neutrinos while Weyl spinors describe massless fermions; flag-dipole spinors \cite{dualtipo4,tipo4epjc,chengflagdipole}, and the recently discovered mass dimension one fermions \cite{mdobook}, which are strong candidates to describe dark matter. Here we analyse in detail the main aspects of each one of the aforementioned type of spinors. 
The very idea of classify things may be accomplished by a given criteria. In the Lounesto's classification \cite{lounestolivro}, spinors are classified according to the physical information encoded on the bilinear forms, quantities that represent the physical observables. Such bilinear forms are restricted by a set of geometrical constraints given by the so-called Fierz-Pauli-Kofink identities. As it will be discussed along this work, the  Lounesto's classification can be divided into two distinct sets: one describing \emph{regular} spinors (single-helicity spinors), and a second one which describes \emph{singular} spinors (dual-helicity spinors), a feature that was not clear or even evidenced until now. We will also show that the regular spinors (also known as Dirac type-1, type-2 and type-3) do not necessarily encompass spinors governed by the Dirac's dynamics, and that any single-helicity belog to the set of regular spinor. 

The concept of helicity plays an important role in modern particle physics. The helicity operator is defined as an inner product between spin projection operator and the operator giving the direction of motion \cite{helicidade}. Our purpose is to unravel how both the helicity \cite{helicidade}, and the way that of connecting the representation spaces, directly influences the spinor dynamics. 

As it is well-defined, all the relativistic fields must satisfy the Klein-Gordon equation, which represents basically the dispersion relation between mass, energy and momentum. Besides that, the Dirac fermions are governed by the Dirac dynamics \cite{ryder}, whereas the Majorana fermions are known to obey the Majorana equation, which is similar to the Dirac one, but also includes the charge conjugate spinor \cite{majoranabook}. On the other hand, particles described by flag-dipole spinors, as well as the mass dimension one spinors, are only governed by the Klein-Gordon equation \cite{mdobook,tipo4epjc,tipo4epjc2}.

We start our analysis by defining both single-helicity and dual-helicity spinors in the spin-$1/2$ representation. After that, we review the main aspects concerning helicity and discrete symmetries, \emph{e.g.} parity operation and charge conjugation operation, and then we will focus in determining the requirements for a given spinor to be an eigenspinor of each aforementioned symmetry operator. Finally, we consider the Lounesto's classification and analyse the fundamental properties of the six classes of spinors.

The paper is organized as follows. Section \ref{prelude} is reserved to define the eigenspinors of the helicity operator. After doing that, we define both  single-helicity and dual-helicity algebraic spinors. Then, we will analyse the interplay between the helicity and the representation spaces, in order to evaluate how such components are connected under the action of parity operator, and also under the Wigner time-reversal operator $\Theta$. Section \ref{capcharge} is reserved to explore the features of the charge conjugation operator $C$. We will study the necessary conditions to define an eigenspinor of the charge conjugation operator, and then we define properly the structure that such spinor must carry to fulfill the requirement of conjugation under $C$. Such a mathematical approach brings an important result. As it will be seen, only single-helicity spinors can be related to parity symmetry, and only dual-helicity spinors can be eigenspinors of charge conjugation operator. In other words, each sector of the Lounesto's classification embrace spinors with a well-defined helicity and, as a consequence, each sector carry eigenspinors of a certain discrete symmetry, P or C. Finally, section \ref{remarks} contains some concluding remarks.


\section{Spinors helicity, parity and Wigner's operators}\label{prelude}
The representation spaces $(0,j)$ and $(j,0)$ are usually connected via parity symmetry $P$, which then leads to the well-known Dirac single-helicity spinors \cite{dirac1928,ryder}. However, if the link between the representation spaces is established through the Wigner time-reversal operator in the spin-$1/2$ representation,
\begin{eqnarray}\label{matrizwigner}
\Theta = \left(\begin{array}{cc}
0 & -1 \\ 
1 & \;\; 0
\end{array}\right), 
\end{eqnarray}
which satisfy the relation \cite{ramond}
\begin{eqnarray}\label{magica}
\Theta\vec{\sigma}\Theta^{-1}=-\vec{\sigma}^*,
\end{eqnarray}
then, we get the Majorana, Elko\footnote{Elko is a German acronym for \emph{Eigenspinoren des
Ladungskonjugationsoperators} which means eigenspinors of charge conjugation operator} and the dual-helicity flag-dipole spinors \cite{ramond,mdobook,tipo4epjc}, which do not satisfy the standard Dirac dynamics. Specifically, we can describe such relations among $(0,1/2)$ and $(1/2,0)$ spaces, as follows
\begin{equation}\label{paritylink}
(1/2,0) \stackrel{P}{\longleftrightarrow} (0,1/2),
\end{equation} 
or
\begin{equation}
(1/2,0) \stackrel{\;\;\;\;\Theta\vec{\sigma}\Theta^{-1}}{\longleftrightarrow}(0,1/2).
\end{equation} 
We can even use the composition of both parity and Wigner time-reversal operator to connect the representation spaces, without any loss of generality. However, once parity is imposed, automatically the Dirac dynamics comes to light (more details can be found in Ref. \cite{diracpauli}). Now, let us consider rotations $\mathcal{R}$, and boosts $\mathcal{B}$, for these representation spaces,
\begin{eqnarray}
\mathcal{R}_{R,L}(\vartheta) &=& e^{i\frac{\vec{\sigma}}{2}\vartheta\hat{n}}\, =\,\cos(\vartheta/2)\mathbbm{1}+ i\vec{\sigma}\hat{n}\sin(\vartheta/2),\label{geradorrotacao}
\end{eqnarray}   
and 
\begin{eqnarray}\label{operadorBboost}
\mathcal{B}_{R,L}(\varphi) &=& e^{\pm\frac{\vec{\sigma}}{2}\varphi\hat{n}}\,=\, \sqrt{\frac{E+m}{2m}}\bigg(\mathbbm{1}\pm\frac{\vec{\sigma}\cdot\vec{p}}{E+m}\bigg),
\end{eqnarray}
where $\vartheta$ represents the angle of rotation, and $\hat{n}$ denote a unit vector along the axis of the rotation. Here, we call the attention to the fact that both the rotation and the boost operators change sign under $P$ or $\mathcal{T}$ operation. Let us now define the momentum in spherical coordinate system, i.e., $p^{\mu} = (E, p\sin\theta\cos\phi, p\sin\theta\sin\phi, p\cos\theta)$. Thus, the helicity operator reads
\begin{eqnarray}\label{helicity}
\vec{\sigma}\cdot \hat{p}=\left(\begin{array}{cc}
\cos\theta & \sin\theta e^{-\textit{i}\phi}\\
\sin\theta e^{\textit{i}\phi} & -\cos\theta
\end{array}
\right),
\end{eqnarray}
where $\vec{\sigma}$ represents the Pauli matrices. From the above parametrization, we can define explictly the rest spinors (in the Weyl representation), i.e., the eigenspinors of the helicity operator,  in the following form\footnote{Actually, we should denote such components just as $\phi^{+}\boldo$ and $\phi^{-}\boldo$. The label \emph{right-hand} and \emph{left-hand} comes from the manner that each one of these components transforms under Lorentz Transformations.} 
\begin{equation}\label{op-helicity}
\vec{\sigma}\cdot \hat{p}\;\phi_{R/L}^{\pm}(k^{\mu}) = \pm\phi_{R/L}^{\pm}(k^{\mu}),
\end{equation}
with
\begin{equation}\label{maodirmais}
\phi_R^{+}(k^{\mu}) = \phi_L^{+}(k^{\mu}) = \sqrt{m}e^{i\vartheta_1}\left(\begin{array}{c}
\cos(\theta/2)e^{-i\phi/2} \\ 
\sin(\theta/2)e^{i\phi/2}
\end{array}\right), 
\end{equation} 
and 
\begin{equation}\label{maodirmenos}
\phi_R^{-}(k^{\mu}) = \phi_L^{-}(k^{\mu}) = \sqrt{m}e^{i\vartheta_2}\left(\begin{array}{c}
\sin(\theta/2)e^{-i\phi/2} \\ 
-\cos(\theta/2)e^{i\phi/2}
\end{array}\right).
\end{equation} 
Here, we have defined
\begin{equation}
k^{\mu}\stackrel{def}{=}\bigg(m,\; \lim_{p\rightarrow 0}\frac{\boldsymbol{p}}{p}\bigg), \; p=|\boldsymbol{p}|.
\end{equation}
As observed in \cite{mdobook}, the presence of the phase factors is necessary to set up the framework of eigenpinors of parity or charge conjugation operators. From eq. (\ref{geradorrotacao}), one can verify that under a rotation by an angle $\vartheta$ the Dirac spinors  pick up a global phase $e^{\pm i\vartheta/2}$, depending on the related helicity. However, this only happens for eigenspinors of parity. For the eigenspinors of charge conjugation operator, the phase factors must be $\vartheta_1=0$ and $\vartheta_2=\pi$, and thus, they are responsible to ensure locality \cite{mdobook}. On the other hand, the mass factor in (\ref{maodirmais}) and (\ref{maodirmenos}) is necessary to ensure that in the massless limit the $(j, 0)\oplus(0, j)$ rest spinors identically vanish, since there cannot be massless particles at rest \cite{majoranalike}. For consistency, the interaction amplitudes must have the factor $m^j$, where the superscript $j$ is the spin of the associated particle \cite{marinov}. It is worth pointing out that the relation $\phi_R^{\pm}(k^{\mu}) = -\phi_L^{\pm}(k^{\mu})$ should not be neglected when dealing with antiparticles, as it has been already noticed and highlighted in \cite{dirac1928,gaioliryder}. 

To define such components in an arbitrary  momentum reference frame, we must perform a Lorentz transformation, 
\begin{equation}
\phi_{R/L}\boldp = e^{\pm\frac{\vec{\sigma}}{2}\varphi\hat{n}}\phi_{R/L}\boldo ,
\end{equation}
in which the boost operator, $e^{\pm\frac{\vec{\sigma}}{2}\varphi\hat{n}}$, was previously defined is Eq. \eqref{operadorBboost}.

Now, we introduce the parity operator $P$,  which inverts the velocity direction, $P\vec{v}\rightarrow -\vec{v}$ and  $P\vec{p}\rightarrow -\vec{p}$. Moreover, the generator of Lorentz boosts changes sign, whereas the generators of rotations keep it unchanged. Thus, by acting with the parity operation on the right-hand and left-hand components, we get
\begin{equation}
P\phi_R\boldp = \sqrt{\frac{E+m}{2m}}\bigg(\mathbbm{1}-\frac{\vec{\sigma}\cdot\vec{p}}{E+m}\bigg)\phi_R\boldo,
\end{equation}
and
\begin{equation}
P\phi_{L}\boldp = \sqrt{\frac{E+m}{2m}}\bigg(\mathbbm{1}+\frac{\vec{\sigma}\cdot\vec{p}}{E+m}\bigg)\phi_L\boldo.
\end{equation}
This is equivalent to
\begin{equation}
P\phi^{\pm}_{R}\boldp \rightarrow  \phi^{\pm}_L\boldp,
\end{equation}
and
\begin{equation}
P\phi^{\pm}_{L}\boldp \rightarrow  \phi^{\pm}_R\boldp,
\end{equation}

which means that parity connects the representation spaces by transforming the right-hand spinor to left-hand one, and \emph{vice-versa}. 

Now, let us analyse the other possibility of connecting the representation spaces, by employing the Wigner time-reversal operator. Bearing in mind the relation (\ref{magica}), we are able to stablish  
\begin{eqnarray}\label{maodirtheta}
[\zeta_1\Theta\phi_{L}^{*}\boldp] &=& \sqrt{\frac{E+m}{2m}}\bigg(\mathbbm{1}+\frac{\vec{\sigma}\cdot\vec{p}}{E+m}\bigg)[\zeta_1\Theta\phi_L^{*}\boldo],
\end{eqnarray}
and 
\begin{eqnarray}\label{maoesqtheta}
[\zeta_2\Theta\phi_{R}^{*}\boldp] &=& \sqrt{\frac{E+m}{2m}}\bigg(\mathbbm{1}-\frac{\vec{\sigma}\cdot\vec{p}}{E+m}\bigg)[\zeta_2\Theta\phi_R^{*}\boldo],
\end{eqnarray}
in which $\zeta_1$ and $\zeta_2$ stands for scalar phases which do not affect the helicity and transformation laws of the right-hand and left-hand components. Now,  we can see that relations (\ref{maodirtheta}) and (\ref{maoesqtheta}) imply that if $\phi_R(p^\mu)$ transforms as $(1/2,0)$, then $[\zeta_2\Theta \,\phi_R^\ast(p^\mu)]$ transforms as $(0,1/2)$ spinor, and similarly if $\phi_L(p^\mu)$ transforms as $(0,1/2)$, then $[\zeta_1\Theta\, \phi_L^\ast(p^\mu)]$ transforms as $(1/2,0)$ spinor, as observed in \cite{ramond}. Now, by applying the parity operation on relations (\ref{maodirtheta}) and (\ref{maoesqtheta}), we get
\begin{equation}
P[\zeta_1\Theta\phi^{\pm *}_{L}\boldp] \rightarrow \phi^{\mp}_R\boldp,
\end{equation}
and 
\begin{equation}
P[\zeta_2\Theta\phi^{\pm *}_{R}\boldp] \rightarrow  \phi^{\mp}_L\boldp.
\end{equation}

We highlight that all the above approaches are considered in a very general framework, where spinors are assumed to satisfy the Klein–Gordon equation and not necessarily the Dirac equation.


\section{Single helicity and dual helicity spinors}
Let us now consider $\psi$ to be a given spinor belonging to a section of the vector bundle $\mathbf{P}_{Spin^{e}_{1,3}}(\mathcal{M})\times_{\rho}\mathbb{C}^4$, where $\rho$ stands for the entire representation space $D^{(1/2,0)}\oplus D^{(0,1/2)}$, or a given sector of such \cite{crawford1,crawford2}. Then, it can be described in the Weyl representation as follows
 \begin{eqnarray}
\psi &=& \left(\begin{array}{c}
\phi_R\\
\phi_L
\end{array}
\right),
\end{eqnarray}
or more explicitly,
\begin{eqnarray}\label{psi}
\psi=\left(\begin{array}{c}
a\\
b\\
c\\
d
\end{array}
\right),
\end{eqnarray}
where the components $a, b, c, d: U\subset \mathbb{R}^{1,3}\to\mathbb{C}$ are scalars fields, and $U$ is an open set of the $\mathbb{R}^{1,3}$. 
Firstly, we consider both spinor components $\phi_{R/L}$ to carry positive helicity, namely
\begin{eqnarray}\label{autohelicity}
\vec{\sigma}\cdot \hat{p}\,\phi_R^{+}=+\phi_R^{+}, \;\;\;\;\mbox{and}\;\;\;\;\; \vec{\sigma}\cdot \hat{p}\,\phi_{L}^{+}=+\phi_{L}^{+},
\end{eqnarray}
leading to the following relations
\begin{eqnarray}\label{b+}
b=\frac{a\sin\theta e^{\textit{i}\phi}}{1+\cos\theta}, \;\;\;\; \mbox{and} \;\;\;\; d=\frac{c\sin\theta e^{i\phi}}{1+\cos\theta}.
\end{eqnarray}
Thus, we can write the first single-helicity spinor as follows,
\begin{eqnarray}\label{positive}
\psi_{(+,+)}=\left(\begin{array}{c}
a\\
\frac{a\sin\theta e^{i\phi}}{1+\cos\theta}\\
c\\
\frac{c\sin\theta e^{i\phi}}{1+\cos\theta}
\end{array}
\right),
\end{eqnarray}
where the labels stand for helicity of the right-hand and left-hand components, respectively. Analogously, for a single-helicity spinor which both components carry negative helicity, restricted to the conditions
\begin{eqnarray}\label{negativehelicity}
\vec{\sigma}\cdot \hat{p}\; \phi_{R}^{-}= -\phi_{R}^{-}, \;\;\;\;\mbox{and}\;\;\;\;\; \vec{\sigma}\cdot \hat{p}\; \phi_{L}^{-}= -\phi_{L}^{-},
\end{eqnarray}
one obtain the following set of relations
\begin{eqnarray}\label{b-}
b=-\frac{a\sin\theta e^{i\phi}}{1-\cos\theta}, \;\;\;\; \mbox{and} \;\;\;\; d=-\frac{c\sin\theta e^{i\phi}}{1-\cos\theta},
\end{eqnarray}
and then, the second single-helicity spinor reads
\begin{eqnarray}\label{shp}
\psi_{(-,-)}=\left(\begin{array}{c}
a\\
\frac{-a\sin\theta e^{i\phi}}{1-\cos\theta}\\
c\\
\frac{-c\sin\theta e^{i\phi}}{1-\cos\theta}
\end{array}
\right).
\end{eqnarray}

Looking towards inspect the action of the parity operator, which may be defined as $\mathcal{P}=m^{-1}\gamma_{\mu}p^{\mu}$ \cite{speranca}, onto the single-helicity spinors \eqref{positive} and \eqref{shp}, we are lead to the following 
\begin{eqnarray}
\mathcal{P}\psi_{(+,+)} =  \left(\begin{array}{c}
c \\ 
\frac{c\sin\theta e^{i\phi}}{1+\cos\theta} \\ 
a \\ 
\frac{a\sin\theta e^{i\phi}}{1+\cos\theta}
\end{array} \right),
\end{eqnarray}
and 
\begin{eqnarray}
\mathcal{P}\psi_{(-,-)} = m \left(\begin{array}{c}
c \\ 
\frac{-c\sin\theta e^{i\phi}}{1-\cos\theta} \\ 
-a \\ 
\frac{a\sin\theta e^{i\phi}}{1-\cos\theta}
\end{array} \right).
\end{eqnarray}
Note that, by imposing $\mathcal{P}\psi_{(+,+)}=+\psi_{(+,+)}$ and $\mathcal{P}\psi_{(-,-)}=-\psi_{(-,-)}$,  we obtain in the first case that $a=c$, and respectively in the second case that $a=-c$, ensuring in such a way the Dirac dynamics. Otherwise, these single-helicity spinors will just obey the Klein-Gordon equation.

Note also that these single-helicity spinors do not necessarily satisfy Dirac dynamics. If we impose the Dirac dynamics, then the parity operation must play the central role connecting the representation spaces \cite{speranca,diracpauli,rodolfosubliminal}, as exemplified above. It is also worth pointing out that another type of single-helicity spinors found in the literature are the Inomata spinors (RIM-spinors)\cite{dinorim}, which are governed by a non-linear equation and classified as regular spinor within Lounesto's classification, agreeing with our results.

On the other hand, we are also able to define a set of dual-helicity spinors by using a very similar procedure. By considering $\vec{\sigma}\cdot \hat{p}\; \phi_{R}^{+}= +\phi_{R}^{+}$ and $\vec{\sigma}\cdot \hat{p}\; \phi_{L}^{-}= -\phi_{L}^{-}$, as well as $\vec{\sigma}\cdot \hat{p}\; \phi_{R}^{-}= -\phi_{R}^{-}$ with $\vec{\sigma}\cdot \hat{p}\; \phi_{L}^{+}= +\phi_{L}^{+}$, we can write the following dual-helicity spinors, 
\begin{eqnarray}\label{dualhel}
\Phi_{(+,-)}=\left(\begin{array}{c}
a\\
\frac{a\sin\theta e^{i\phi}}{1+\cos\theta}\\
c\\
\frac{-c\sin\theta e^{i\phi}}{1-\cos\theta}
\end{array}
\right), \qquad
\Phi_{(-,+)}=\left(\begin{array}{c}
a\\
\frac{-a\sin\theta e^{i\phi}}{1-\cos\theta}\\
c\\
\frac{c\sin\theta e^{i\phi}}{1+\cos\theta}
\end{array}
\right).
\end{eqnarray}
These dual-helicity spinors are in very agreement with the singular spinors structure previously defined in \cite{cavalcanticlassification}, which can be written as 
\begin{eqnarray}\label{cavl}
\Phi_{h}=\left(\begin{array}{c}
\frac{-bcd^{*}}{|c|^{2}}\\
b\\
c\\
d
\end{array}
\right),
\end{eqnarray}
where the index $h$ stands for the helicity. We notice that these dual-helicity spinors have the same form of the flag-pole and flag-dipole spinors in Ref \cite{cavalcanticlassification}. The $\Phi_{(\pm,\mp)}$ spinor components are clearly related by the $\Theta$ operator, which is responsible to connect the $(0,1/2)$ and $(1/2,0)$ representation spaces. Moreover, a remarkably feature concerning dual-helicity spinors is that they do not satisfy the Dirac dynamics, i.e., dual-helicity spinors do not form a set of eigenspinors of parity operator. To illustrate the above statements, we may act with the Dirac operator, $\gamma_{\mu}p^{\mu}$, on the dual-helicity spinors (\ref{dualhel}), obtaining the following set of relations 
\begin{eqnarray}
\gamma_{\mu}p^{\mu}\Phi_{(+,-)} \propto \Phi_{(-,+)}, \label{diracdual1}
\\
 \gamma_{\mu}p^{\mu}\Phi_{(-,+)} \propto \Phi_{(+,-)}. \label{diracdual2}
\end{eqnarray}
The above equations translates into $\mathcal{P}\Phi_{h} \propto \Phi_{h^{\prime}}$, making explicit that the action of the parity operator over dual-helicity spinors flips the helicity label. Due the dual-helicity feature, such spinors  hold \emph{degeneracy} under spatial reflection, and then preventing them from forming a set of eigenspinors of parity operator \cite{elkostates}.

Therefore, above results lead us to conclude that the singular spinors --- flag-pole and also flag-dipole spinors --- carry dual-helicity properties, evincing the fact that, besides single-helicity (regular spinors), the Lounesto's classification also embraces dual-helicity spinors (singular spinors). This is the first time that this issue is reported, and it is an important outcome of this work, corroborating with the previous statements in \cite{cavalcanti4,cavalcanticlassification}. Such results are extremely important. Thus, after all the previous calculations presented, we have shown that the Lounesto's classification present a strong dichotomy, since spinors can be divided in two distinct sectors: one sector embracing single-helicity spinors, and another composed by dual-helicity spinors, namely 
\begin{eqnarray}
&&\left. \begin{array}{rllll}
1. & \sigma\neq0, & \omega\neq0, &\boldsymbol{K}\neq0, & \boldsymbol{S}\neq0, \\
2. & \sigma\neq0, &\omega=0, &\boldsymbol{K}\neq0, & \boldsymbol{S}\neq0, \\
3. & \sigma=0, & \omega\neq0, &\boldsymbol{K}\neq0, & \boldsymbol{S}\neq0,
\end{array}\right\}\mbox{\emph{single-helicity}}. 
\nonumber\\
&&\left. \begin{array}{rlll}
4. & \sigma=0=\omega, & \boldsymbol{K}\neq0, & \boldsymbol{S}\neq0, \\
5. & \sigma=0=\omega, & \boldsymbol{K}=0, & \boldsymbol{S}\neq0,  \\
\end{array}\right\}\mbox{\emph{dual-helicity}}.
\\\nonumber
&&\left. \begin{array}{rlll}
6. & \sigma=0=\omega, & \boldsymbol{K}\neq0, & \boldsymbol{S}=0.
\end{array}\right\}\mbox{\emph{single-helicity}}.
\end{eqnarray}

We have labelled Weyl spinor as ``\emph{single-helicity}'', since such spinors always carry a null component \cite{cavalcanticlassification,rodolfoconstraints,rodolfosubliminal}, \textit{i.e.},
\begin{eqnarray}
\psi_{6_{1}} = \left(\begin{array}{c}
\phi_R \\ 
0
\end{array}\right) \quad\mbox{and} \quad  \psi_{6_{2}} = \left(\begin{array}{c}
0 \\ 
\phi_L
\end{array}\right).
\end{eqnarray}
In this vein, we believe that such feature allows us to define properly only one spinor's component helicity, thus standing for a single-helicity spinor.


\section{Charge Conjugation operator}\label{capcharge}

In this section, we will explore some important properties concerning the charge conjugation operation. Let us consider the case when a given spinor is an eigenspinor of charge conjugation operator, which is defined as follows \cite{mdobook}   
\begin{eqnarray}\label{C}
C=\left(\begin{array}{cc}
\mathbb{O} & i\Theta\\
-i\Theta & \mathbb{O}
\end{array}
\right)\mathcal{K},
\end{eqnarray}
where $\mathbb{O}$ stand for a $2\times 2$ null-matrix, $\Theta$ stand for the Wigner time-reversal operator given in (\ref{matrizwigner}),  and $\mathcal{K}$ stand for the algebraic complex conjugation operation. 
Regarding the structure of the eigenspinors of the charge conjugation operator, some questions naturally arise. For instance, we wonder if a given eigenspinor of charge conjugation operator is also  an eigenspinor of the parity operator or vice versa?. Also, if such symmetry imposes any constraint on the helicity?. And if that symmetry depends on the way that the representation spaces are connected?. 

First of all, we will focus our attention on the spinors described by the equation (\ref{psi}). If we demand that such spinors are eigenspinors of $C$, namely
\begin{eqnarray}\label{conjugacao}
C\Phi^{S/A}=\pm\Phi^{S/A},
\end{eqnarray}
where the upper index stand for \emph{self-conjugated} and \emph{anti self-conjugated}, respectively, then the spinor components must satisfy following relations
\begin{eqnarray}\label{compo1}
a=-id^{*},\;\;\;\; b=ic^{*},
\end{eqnarray}
with
\begin{eqnarray}\label{compo2}
||a||^{2}=||d||^{2},\;\;\;\; ||b||^{2}=||c||^{2}.
\end{eqnarray}
These constraints  determine uniquely the eigenspinors of charge conjugation operator. In fact, if we impose  \mbox{$C\Phi^{S} = +\Phi^{S}$}, we reach to the following spinor 
\begin{eqnarray}\label{Cpsi+}
\Phi^S=\left(\begin{array}{c}
-id^{*}\\
ic^{*}\\
c\\
d
\end{array}
\right).
\end{eqnarray}
And, by imposing $C\Phi^{A} = -\Phi^{A}$, we obtain
\begin{eqnarray}\label{Cpsi-}
\Phi^{A}=\left(\begin{array}{c}
id^{*}\\
-ic^{*}\\
c\\
d
\end{array}
\right).
\end{eqnarray}
From  equations (\ref{Cpsi+}) and (\ref{Cpsi-}), one is able to define the connection between dual-helicity spinors and the charge conjugation operator, and also to claim that only dual-helicity spinors which have the representation spaces connected via $\Theta$ operator may be eigenspinors of charge conjugation operator. A concrete counterexample of this result lies in the Elko dual-helicity spinors \cite{mdobook} (see also \cite{dualtipo4,chengflagdipole}). Note that, Dirac spinors do not have the structure given in \eqref{Cpsi+} and \eqref{Cpsi-}, and then they can never satisfy \eqref{conjugacao}. Although the dynamics in each class of Lounesto’s classification is still an open issue, this fact shows us that there might be more types of spinors with unknown dynamics in the Lounesto's classification \cite{bonorapandora}.

A direct analysis of the spinors given in \eqref{Cpsi+} and \eqref{Cpsi-} lead to conclude that dual-helicity spinors stands for class of spinors holding strong candidates of eigenspinor of the charge conjugation operator.  Nonetheless, just a (really) small subset of singular spinors may accomplish such a task --- class 5 --- thus, only flag-pole spinors, as the Elko and Majorana spinors, hold conjugacy under $C$ \cite{cavalcanticlassification,rodolfosubliminal,hierarchy}.

\subsection{Time-reversal operator and its consequences}
Having examined some important properties of parity operator ($\mathcal{P}=m^{-1}\gamma_{\mu}p^{\mu}$) and the charge conjugation operator, both acting under single-helicity and dual-helicity spinors, let us now inspect  some consequences of the time-reversal operator $\mathcal{T}=i\gamma_5C$, where $\gamma_5=-i\gamma_{0123}$, and $C$ stand for the charge-conjugation operator. Let us start by defining the action of the time-reversal operator on a general spinor \eqref{psi}, as follows
\begin{eqnarray}
\mathcal{T}\left(\begin{array}{c}
a \\ 
b \\ 
c \\ 
d
\end{array} \right) = \left(\begin{array}{c}
d^* \\ 
-c^* \\ 
b^* \\ 
-a^*
\end{array} \right).
\end{eqnarray}  
In order to obtain an eigenspinor of time-reversal operator it is necessary to satisfy the following conditions
\begin{eqnarray}
&&a = d^*, \label{T-1}
\\
&&b=-c^*,
\\
&&c = b^*, 
\\
&&d=-a^*. \label{T-4}
\end{eqnarray}
However, it is clear that the above conditions are incongruous relations, and thus making impossible to define an eigenspinor of the time-reversal operator. We highlight that even acting with $\mathcal{T}$ over the single-helicity spinors or a dual-helicity spinor, defined in \eqref{positive}, \eqref{shp}, \eqref{Cpsi+} and \eqref{Cpsi-}, the relations \eqref{T-1}-\eqref{T-4} hold true, yielding for the singçe-helicity spinors
\begin{eqnarray}
\mathcal{T}\psi_{(+,+)} = \left(\begin{array}{c}
\frac{c^*\sin\theta e^{-i\phi}}{1+cos\theta} \\ 
-c^* \\ 
\frac{a^*\sin\theta e^{-i\phi}}{1+cos\theta} \\ 
-a^*
\end{array} \right),\quad \mathcal{T}\psi_{(-,-)} = \left(\begin{array}{c}
\frac{-c^*\sin\theta e^{-i\phi}}{1-cos\theta} \\ 
-c^* \\ 
\frac{-a^*\sin\theta e^{-i\phi}}{1-cos\theta} \\ 
-a^*
\end{array} \right), 
\end{eqnarray}
and for the dual-helicity spinors, it furnishes
\begin{eqnarray}
\mathcal{T}\Phi_{(+,-)} = \left(\begin{array}{c}
\frac{-c^*\sin\theta e^{-i\phi}}{1-cos\theta} \\ 
-c^* \\ 
\frac{a^*\sin\theta e^{-i\phi}}{1+cos\theta} \\ 
-a^*
\end{array} \right), \quad \mathcal{T}\Phi_{(-,+)} = \left(\begin{array}{c}
\frac{-c^*\sin\theta e^{-i\phi}}{1+cos\theta} \\ 
-c^* \\ 
\frac{-a^*\sin\theta e^{-i\phi}}{1-cos\theta} \\ 
-a^*
\end{array} \right).
\end{eqnarray}
The outcome of this section, is that time-reversal operator is not able to connect the $(0, 1/2)$ and $(1/2, 0)$ representation spaces.


\section{Final Remarks}\label{remarks}
In this work, we have explored the underlying concepts concerning the spinorial structure, discrete symmetries, and their relations with the representation space $(1/2,0)\oplus(0, 1/2)$. From the very definition of the eigenspinors of the helicity operator, and the way that right-hand and left-hand components relate to each other, we have constructed single-helicity spinors which do not necessarily obey the Dirac equation. However, they are classified as Dirac spinors within the Lounesto's classification. Nonetheless, when parity is introduced  to connect both sectors of the representation space, the Dirac dynamics is automatically reached, and thus, a specific class of spinor within the Lounesto's classification satisfying Dirac dynamics is obtained. Moreover, we have explicitly shown that single-helicity spinors do not satisfy all the necessary requirements to compose a set of eigenspinors of the charge conjugation operator.

On the other hand, the dual-helicity spinors must be necessarily defined through the Wigner time-reversal operator, which plays the role of linking the representation spaces. Such feature ensures the conjugacy under the $C$ symmetry, as it was shown in equations \eqref{cavl}, \eqref{Cpsi+} and \eqref{Cpsi-}. However, due to the fact that it does not carry the parity as a link between the representation spaces, it cannot be an eigenspinor of parity operator, and consequently, dual-helicity spinors do not obey the Dirac dynamics. Besides that, not  every dual-helicity spinor must necessarily be an eigenspinor of charge conjugation operator, as it is the case of the  Elko spinors \cite{mdobook}, and also of the recently proposed dual-helicity flag-dipole spinors \cite{dualtipo4,tipo4epjc,tipo4epjc2}.

For completeness, we have analysed some aspects of the regular and singular spinors within Lounesto's classification. We highlight that regular spinors (Lounesto's classes 1, 2 and 3) stands for the single-helicity ones. Regarding the remaining classes, except for class-6, we point out that singular spinors are endowed with dual-helicity feature, being strong candidates to carry intrinsic \emph{darkness}\cite{aaca,dharamnewfermions, rodolfompla,tipo4epjc}. Then, the importance of our results lies on the fact that, unlike it was believed in the literature, the Lounesto's classification do not embrace exclusively single-helicity spinors, but also dual-helicity spinors.\\


\section{Acknowledgements}
CHCV thanks CNPq PCI Grant N$^{\circ}$. 300236/2019-0 for the financial support and ARA would like to thank CAPES-Brazil for partial financial support.

\bibliographystyle{unsrt}
\bibliography{refs}

\begin{thebibliography}{10}

\bibitem{dualtipo4}
R.J.~Bueno Rogerio and C.H.~Coronado Villalobos.
\newblock Some remarks on dual helicity flag-dipole spinors.
\newblock {\em Physics Letters A}, 383(30):125873, 2019.

\bibitem{tipo4epjc}
R.~J.~Bueno Rogerio, C.~H.~Coronado Villalobos, and A.~R. Aguirre.
\newblock A hint towards mass dimension one {F}lag-dipole spinors.
\newblock {\em The European Physical Journal C}, 79(12):991, 2019.

\bibitem{chengflagdipole}
C.~Y. Lee.
\newblock {Mass dimension one fermions from flag dipole spinors}.
\newblock 2018.

\bibitem{mdobook}
D.~Ahluwalia.
\newblock {\em Mass Dimension One Fermions}.
\newblock Cambridge University Press, 2019.

\bibitem{lounestolivro}
P.~Lounesto.
\newblock {\em Clifford algebras and spinors}, volume 286.
\newblock Cambridge university press, 2001.

\bibitem{helicidade}
C.~G. Boehmer and L.~Corpe.
\newblock Helicity—from {C}lifford to graphene.
\newblock {\em Journal of Physics A: Mathematical and Theoretical},
  45(20):205206, 2012.

\bibitem{ryder}
L.~H. Ryder.
\newblock {\em Quantum Field Theory}.
\newblock Cambridge University Press, 2 edition, 1996.

\bibitem{majoranabook}
A.~Franklin.
\newblock {\em Are there really neutrinos? {A}n evidential history}.
\newblock CRC Press, 2018.

\bibitem{tipo4epjc2}
R.~J.~Bueno Rogerio, A.~R. Aguirre, and C.~H.~Coronado Villalobos.
\newblock Flag-dipole spinors: On the dual structure derivation and
  $\mathcal{C}$, $\mathcal{P}$ and $\mathcal{T}$ symmetries, 2020.

\bibitem{dirac1928}
P.~A.~M. Dirac.
\newblock The quantum theory of the electron.
\newblock {\em Proceedings of the Royal Society of London. Series A, Containing
  Papers of a Mathematical and Physical Character}, 117(778):610--624, 1928.

\bibitem{ramond}
P.~Ramond.
\newblock Field theory: {A} modern primer.
\newblock {\em Front. Phys.}, 74:1--397, 1981.

\bibitem{diracpauli}
C.~H. Coronado~Villalobos and R.~J. Bueno~Rogerio.
\newblock The connection between {D}irac dynamic and parity symmetry.
\newblock {\em EPL (Europhysics Letters)}, 116(6):60007, 2016.

\bibitem{majoranalike}
D.~V. Ahluwalia and M.~B. Johnson.
\newblock Majorana-like representation spaces: Construction and physical
  interpretation.
\newblock {\em arXiv preprint hep-th/9307118}, 1994.

\bibitem{marinov}
M.~S. Marinov.
\newblock Construction of invariant amplitudes for interactions of particles
  with any spin.
\newblock {\em Annals of Physics}, 49(3):357--392, 1968.

\bibitem{gaioliryder}
F.~H. Gaioli and E.~T. Garcia~Alvarez.
\newblock Some remarks about intrinsic parity in {R}yder’s derivation of the
  {D}irac equation.
\newblock {\em American Journal of Physics}, 63(2):177--178, 1995.

\bibitem{crawford1}
J.~P. Crawford.
\newblock Bispinor geometry for even-dimensional space-time.
\newblock {\em Journal of mathematical physics}, 31(8):1991--1997, 1990.

\bibitem{crawford2}
J.~P. Crawford.
\newblock Clifford algebra: Notes on the spinor metric and {L}orentz,
  {P}oincar{\'e}, and conformal groups.
\newblock {\em Journal of mathematical physics}, 32(3):576--583, 1991.

\bibitem{speranca}
L.~D. Sperança.
\newblock An identification of the {D}irac operator with the parity operator.
\newblock {\em International Journal of Modern Physics D}, 23(14):1444003,
  2014.

\bibitem{rodolfosubliminal}
R.~J.~Bueno Rogerio.
\newblock Subliminal aspects concerning the {L}ounesto’s classification.
\newblock {\em The European Physical Journal C}, 80(4), 2020.

\bibitem{dinorim}
D.~Beghetto and J.~M. Hoff~da Silva.
\newblock The (restricted) {I}nomata-{M}ckinley spinor representation and the
  underlying topology.
\newblock {\em EPL (Europhysics Letters)}, 119(4):40006, 2017.

\bibitem{cavalcanticlassification}
R.~T. Cavalcanti.
\newblock Classification of singular spinor fields and other mass dimension one
  fermions.
\newblock {\em International Journal of Modern Physics D}, 23(14):1444002,
  2014.

\bibitem{elkostates}
J.~M. Hoff~da Silva and R.~J. Bueno~Rogerio.
\newblock Massive spin-one-half one-particle states for the mass-dimension-one
  fermions.
\newblock {\em EPL (Europhysics Letters)}, 128(1):11002, 2019.

\bibitem{cavalcanti4}
R.~T. Cavalcanti.
\newblock Looking for the classification of singular spinor fields dynamics and
  other mass dimension one fermions: Characterization of spinor fields.
\newblock {\em International Journal of Modern Physics D}, 23, 07 2014.

\bibitem{rodolfoconstraints}
R.~J.~Bueno Rogerio.
\newblock Constraints on mapping the {L}ounesto’s classes.
\newblock {\em The European Physical Journal C}, 79(11):929, 2019.

\bibitem{bonorapandora}
L.~Bonora, J.~M.~Hoff da~Silva, and R.~da~Rocha.
\newblock Opening the {P}andora's box of quantum spinor fields.
\newblock {\em The European Physical Journal C}, 78(2):157, 2018.

\bibitem{hierarchy}
R.~J.~Bueno Rogerio.
\newblock Singular spinors and their connection.
\newblock {\em arXiv:2003.11368}, 2020.

\bibitem{aaca}
D.~V. Ahluwalia.
\newblock The theory of local mass dimension one fermions of spin one half.
\newblock {\em Advances in Applied Clifford Algebras}, 27(3):2247--2285, Sep
  2017.

\bibitem{dharamnewfermions}
D.~V. Ahluwalia.
\newblock A new class of mass dimension one fermions.
\newblock {\em Proceedings of the Royal Society A}, 476(2240):20200249, 2020.

\bibitem{rodolfompla}
R.~J.~Bueno Rogerio.
\newblock From dipole spinors to a new class of mass dimension one fermions.
\newblock {\em Modern Physics Letters A}, page 2050319, 2020.

\end{thebibliography}

\end{document}